\documentclass[doublecol]{epl2}
\usepackage{graphicx}
\usepackage{amsmath,amssymb,revsymb,epsfig}

\title{Impact of link deletions on public cooperation in scale-free\\
networks}
\shorttitle{Impact of link deletions on public cooperation in scale-free networks}

\author{Luo-Luo Jiang,\inst{1,2,3} Matja{\v z} Perc,\inst{4} Wen-Xu Wang,\inst{1} Ying-Cheng Lai,\inst{1,5} Bing-Hong Wang\inst{3,6,7}}
\shortauthor{Luo-Luo Jiang \textit{et al.}}

\institute{\inst{1}School of Electrical, Computer and Energy Engineering, Arizona State University, Tempe, Arizona, USA\\
\inst{2}College of Physics and Technology, Guangxi Normal University, Guilin, Guangxi, China\\
\inst{3}Department of Modern Physics, University of Science and Technology of China, Hefei, China\\
\inst{4}Department of Physics, Faculty of Natural Sciences and Mathematics, University of Maribor, Slovenia, EU\\
\inst{5}Department of Physics, Arizona State University, Tempe, Arizona, USA\\
\inst{6}Research Center for Complex System Science, University of Shanghai for Science and Technology, Shanghai, China\\
\inst{7}Shanghai Academy of System Science, Shanghai, China}

\abstract{Working together in groups may be beneficial if compared to isolated efforts. Yet this is true only if all group members contribute to the success. If not, group efforts may act detrimentally on the fitness of their members. Here we study the evolution of cooperation in public goods games on scale-free networks that are subject to deletion of links that are connected to the highest-degree individuals, i.e., on networks that are under attack. We focus on the case where all groups a player belongs to are considered for the determination of payoffs; the so-called multi-group public goods games. We find that the effect of link deletions on the evolution of cooperation is predominantly detrimental, although there exist regions of the multiplication factor where the existence of an ``optimal'' number of removed links for the \textit{deterioration} of cooperation can also be demonstrated. The findings are explained by means of wealth distributions and analytical approximations, confirming that socially diverse states are crucial for the successful evolution of cooperation.}

\pacs{02.50.Le}{Decision theory and game theory}
\pacs{89.75.Hc}{Networks and genealogical trees}
\pacs{87.23.Ge}{Dynamics of social systems}

\begin{document}
\maketitle

\section{Introduction}

Understanding the evolution of cooperation is a fundamental problem in social and biological sciences \cite{axelrod84, nowakXs06}. As a common theoretical framework, the prisoner's dilemma game \cite{hofbauer88} has received ample attention in the study of cooperation between selfish individuals via pairwise interactions. However, cooperation is observed not only in pairwise interactions, but indeed even more so in groups involving more than two individuals. Examples are many, including public
transportation \cite{helbingXrmp01} and attempts of averting
global warming \cite{milinskiXpnas08}. Public goods games (PGGs)
are established, both theoretically as well as experimentally, to
capture the essence of the dilemma that underlies such cooperative behavior. In a typical PGG experiment, $n$ players are asked to invest into a common pool. They all know that the accumulated contributions will be multiplied by a factor $r>1$, and then
equally divided among all participants irrespective of their
investment. If all players cooperate they increase their initial input by $(r-1)c$, where $c$ is the invested contribution, i.e., the cost of cooperation. However, every player is faced with the temptation to defect by exploiting the
contributions of other players and withholding its own. In particular,
if a player chooses to defect while others cooperate it will get a
higher net payoff. It is thus more likely that this strategy will spread
in an environment governed by natural selection due to the higher fitness of defectors. The conflict escalates as more and more players adopt the defecting
strategy. In doing so, they neglect collective interests, which
ultimately leads to social downfall and poverty, known also as the ``tragedy of the commons'' \cite{hardin_g_s68}. Human experiments reveal, however, that the level of cooperation is surprisingly high in one round games, but deceases fast from round
to round if a game is played repeatedly \cite{fehrXaer00}.

Thus, in PGGs individuals face the temptation to reap benefits on the expense of others, and in doing so likely induce further defecting actions which harm collective well-being. In contrast, cooperators are able to resist these temptations with
the aim of facilitating group benefits and social welfare. Defectors therefore bear
no costs when collecting identical benefits as cooperators, which
ultimately leads to widespread defection. This theoretical
prediction, however, disagrees with experimental findings
\cite{fehrXars07}. To resolve this disagreement between
observations and theory, several mechanisms have been proposed
that promote cooperation. Hauert \textit{et al.} \cite{hauertXs02} introduced voluntary
participation in PGGs and found that it supports
cooperation in a Red-Queen-like manner. Szab\'{o} \textit{et al.}
\cite{szaboXprl02} also studied the impact of voluntary
participation in PGGs on a square lattice, reporting
that the introduction of loners leads to a cyclic dominance of the
three strategies. Punishment too has been identified as a
viable route to cooperative behavior \cite{fehrXn02, boydXpnas03,
hauertXpnas06}, although its effectiveness depends on whether the
participation in the PGG is optional \cite{hauertXs07}, and on whether the interactions amongst players are structured \cite{helbing_ploscb10}. The reward as a means to cooperation in spatial public goods games has also been considered \cite{szolnoki_epl10}. Recently, social diversity introduced via heterogeneous interaction networks and participation of players in multiple groups \cite{santosXn08, YWWLW:2009} was identified as a viable route to cooperative behavior in PGGs, as was the
relaxation of strategy adoption criteria to account for mutation
and random exploration of available strategies \cite{traulsenXpnas09}. Effects of different temporal and spatial scales in evolutionary games have also been revealed as being potentially important \cite{rocaXprl06, wuzxXpre09,JWLWpre10}.
Fowler and Christakis observed experimentally that cooperative
behavior can spread among people with personal ties \cite{FCpnas10}. However, quantitative research on the effects of altering the social ties in multi-group PGGs is more difficult to come by \cite{segbroeckXprl09}.

In this paper, we therefore study the evolution of cooperation in PGGs
on scale-free networks \cite{albertXrmp02, newmanXsiam03, boccalettiXpr06} where players can contribute and receive payoffs from all the groups with which they are affiliated. More precisely, in a multi-group PGG, every individual
participates also in the groups centered around its direct neighbors, as was
proposed recently in \cite{santosXn08}. The topology of the
interaction network also plays an important role in the evolution
of cooperation (for a comprehensive review see \cite{szaboXpr07}). In particular, it is well-known that heterogeneous interaction networks promote cooperation if the interactions are pairwise \cite{santosXprl05, poncelaXepl09}. Since, in a scale-free network, there exists a small set of nodes with many more links attached to them than to other nodes forming the network, it is important to assess the role of this highly connected nodes by the evolution of cooperation. We are thus led to consider multi-group PGGs on scale-free networks that are subject to deletion of links connected to the highest degree individuals.

We shall examine quantitatively the effects of network
heterogeneity on cooperation in multi-group PGGs. In particular, we generate scale-free networks according to the method in \cite{barabasiXs99} and then remove a number of links connected to the highest degree nodes. The deletion of such links is traditionally referred to as network attack \cite{albertXn00}, and it was shown that the resilience to
such actions is crucial for situations such as fast information transfer within the
world-wide-web \cite{pastorXprl01}, uninterrupted supply with electricity \cite{albertXpre04}, fast spread of epidemics and viral infections \cite{pastorXpre02, zanetteXpa02, barthelemyXprl04, colizzaXplos07}, and cascade failure in networks \cite{motterXpre02}. Several studies elaborated analytically on the resilience of complex networks via the usage of percolation theory \cite{cohenXprl00, callawayXprl00, cohenXprl01, pietschXpre06}. Interestingly, we find that the deletions mainly impair the evolution of cooperation in multi-group PGGs. However, the effect is monotonously negative only within a rather narrow region of the multiplication factor, whereas outside this region there exists an ``optimal'' number of deleted links at which the fraction of cooperators on the network is minimal. We examine the resulting distributions of wealth in terms of the dependence on the number of removed links in order to shed light on the reported results. We confirm that socially diverse states are crucial for the successful evolution of cooperation, as was argued recently in \cite{santosXn08}.

The remainder of this paper is organized as follows. In the next section we describe the employed multi-group PGG, and in Section III we present numerical results. Section IV
features an analytical treatment that provides further insights and explains our findings. Lastly, we summarize the main results and outline their potential implications.

\section{Multi-group public goods games on scale-free networks}

\begin{figure}
\centerline{\epsfig{file=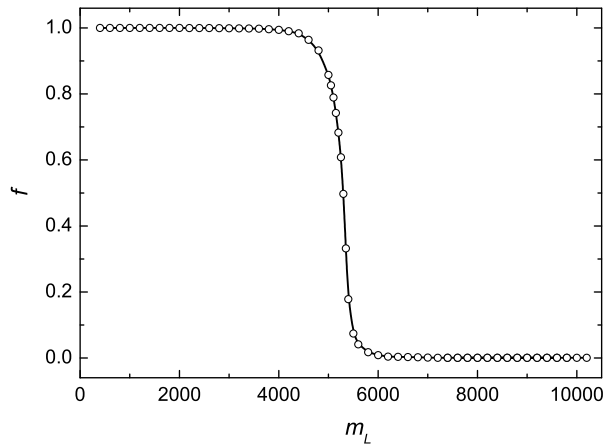,width=8.0cm}}
\caption{Fraction of the largest connected component $f$ in dependence on the number of deleted links $m_{L}$. The original network size here was $N=5000$, and the results were obtained as averages over $1000$ different initial network configurations.}
\label{networks}
\end{figure}

In reality, interactions among individuals can often be described by complex networks. Here we focus on scale-free networks whereby each vertex corresponds to a particular player $x$. The generation of the network starts with two connected players, and subsequently every new player is attached to two old players already present in the network, whereby the probability $\Pi$ that a new player will be connected to an old player $x$ depends on its degree $k_x$ according to $\Pi=k_x / \sum k_y$. This growth and preferential attachment scheme yields a network with an average degree $\kappa=(1/N) \sum k_x$ of four, and a power-law degree distribution with the degree exponent of $3$ \cite{barabasiXs99}. Before the PGG starts, we delete a total of $m_{L}$ links
that are connected to the highest degree nodes. According to Albert \textit{et al.} \cite{albertXn00}, this corresponds to an attack, and we are interested in how different values of $m_{L}$ affect the evolution of cooperation in multi-group PGGs. As shown in Fig.~\ref{networks}, with increasing of $m_{L}$ there exists a transition from an
integrated network (there exists a path between any pair of nodes) to a fragmented network, where some groups of nodes become isolated and cannot be reach from the largest network component. The fraction $f=S/N$ can then be defined, where $S$ is the size of
the largest connected component and $N$ is the size of the original network (here we have $N=5000$). Note that for $m_{L} < 4000$, $S \approx N$ so that the original network remains essentially intact. For $m_{L}>6000$, however, $S \ll N$. At the transition point the value of $m_{L}$ is about $5000$ if the size of the network is $N=5000$, as was the case in Fig.~\ref{networks}.

Initially each player on site $x$ is designated either as a cooperator ($s_x = C$) or as a defector ($s_x = D$) with equal probability. Cooperators contribute $c$ (cost of cooperation) to each public goods game, while defectors contribute nothing. Subsequently, the total contribution that accumulates within a group is multiplied by the multiplication factor $r>1$ that reflects synergistic effects of cooperation, and the resulting amount is divided equally among all the players irrespective of their strategies (and initial contributions).

\begin{figure}
\centerline{\epsfig{file=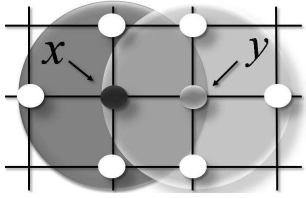,width=4.0cm}}
\caption{Schematic presentation of a multi-group public goods game. Player $x$ is not only a member of the group in which it is focal (the group is marked with a dark gray circle), but it is also a member of four other groups (one such group centered around its neighbor $y$ is marked with a light gray circle.)}
\label{group}
\end{figure}

Since players participate in multiple groups, each player $x$ acquires its total payoff $P_x$ not only by participating in the group consisting of its direct neighbors, but also in all the groups that are centered around its direct neighbors (see Fig.~\ref{group}) \cite{santosXn08}. Figure~\ref{group} demonstrates such a setup schematically, where player $x$ is a member of five groups, having five members each. One is the group centered around player $x$ (marked with a dark gray circle), but there are also four other groups centered around its neighboring players in which player $x$ is also member. An example of such a group is marked with a light gray circle and is centered around player $y$. The payoff of player $x$ with strategy $s_x$ that is associated with the group centered around player $y$ is then
\begin{equation}
p_{x,y}=\frac{r}{k_y+1}\sum_{z=0}^{k_y}\frac{c}{k_z+1}s_z-\frac{c}{k_x+1}s_x,
\end{equation}
where $z=0$ stands for $y$, $s_z$ is the strategy of the neighbor
$z$ of $y$, and $k_z$ is its degree. The strategy is $s_x=1$ if $x$ cooperates and $s_x=0$ if it defects. The cost $c$ refers to the contribution of every cooperator in each group it is a member. For example, if a cooperator is member in five groups, then its total cost will be $5c$, i.e., one $c$ for each group (or equivalently, one $c$ for each public goods game played. Without loss of generality, the cost $c$ is set equal to $1$. The total payoff of player $x$ comes from the summation over all the groups
\begin{equation}
P_x=\sum_{y \in \Gamma_x}p_{x,y},
\end{equation}
where $\Gamma_x$ denotes the groups that are centered around the neighbors $y$ and the player $x$ itself. The number of groups in which player $x$ participates is thus $k_{x}+1$.

At each time step $t$, each player $x$ acquires its payoff $P_x$ as described above. Subsequently, all players update their strategy synchronously by selecting at random one of their direct neighbors $y$. If $P_x > P_y$ player $x$ keeps its strategy
$s_x$, but if $P_x < P_y$ player $x$ adopts the strategy of player $y$ with the probability
\begin{equation}
W(s_y \to s_x)=\frac{P_{y}-P_{x}}{\Delta P_{max}},
\end{equation}
where $\Delta P_{max}$ ensures the proper normalization and is given by the maximal possible payoff difference between players $x$ and $y$ \cite{santosXn08}. Results presented below are obtained on networks hosting $N=5000$ players. Equilibrium
fractions of cooperators $\rho_{C}$ are determined within $5000$ time steps after sufficiently long transients were discarded.

\section{Simulation results}

\begin{figure}
\centerline{\epsfig{file=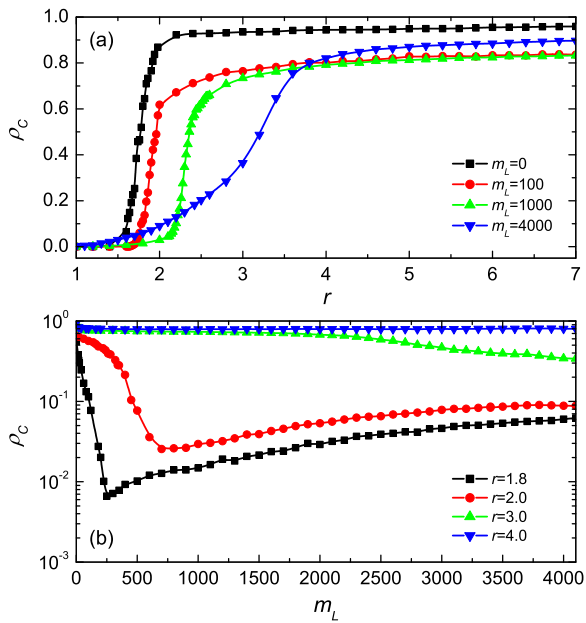,width=7.8cm}}
\caption{(a) Fraction of cooperators $\rho_{C}$ in dependence on the multiplication factor $r$ for different numbers $m_L$ of deleted links connected to the highest degree players. (b) Fraction of cooperators $\rho_{C}$ in dependence on $m_L$ for different multiplication factors $r$. Note that for certain values of $r$ there exists an ``optimal'' number of deleted links $m_L = m_{L}^{opt}$, at which defection thrives best. This effect, however, is quite subtle (see main text for details). The vertical axis has a logarithmic scale.}
\label{reward}
\end{figure}

First, we investigate the dependence of the fraction of cooperators $\rho_C$
on the enhancement factor $r$ for different numbers of removed links attached to the main hubs $m_L$, as shown in Fig.~\ref{reward}(a). It can be observed that the evolution of
cooperation due to link deletions is affected strongly in multi-group PGGs. In particular, the cooperation at a given value of $r$ as well as the emergence of complete cooperator dominance both vary significantly in their dependence on $m_L$. Indeed, if
$m_L=4000$ links are removed from the network before the multi-group PGG starts, the evolution of cooperation will be greatly impaired practically across the whole span of $r$, especially if compared to the $m_L=0$ case. We can thus conclude that the
decrease in network heterogeneity due to link deletions acts predominantly detrimental on the evolution of cooperation in multi-group PGGs.

\begin{figure}
\centerline{\epsfig{file=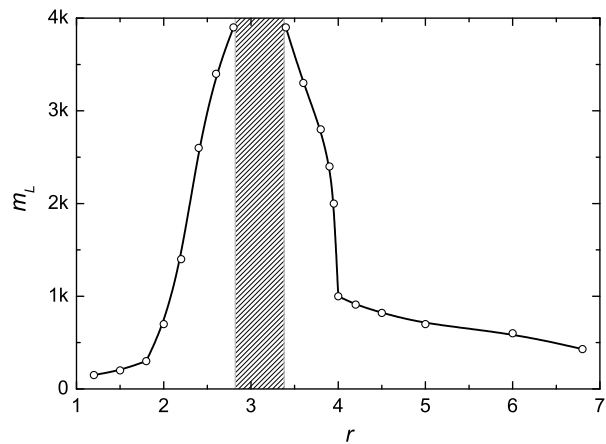,width=8.0cm}}
\caption{The existence and dependence of the ``optimal'' number of deleted links $m_{L}^{opt}$, where defection thrives best, in dependence on the multiplication factor $r$. There exists only a rather narrow region (hatched gray) where the impact of link deletions is monotonously negative ($m_{L}^{opt}$ does not exist). Outside this region, however, there always exists $m_{L}^{opt}$ at which $\rho_{C}$ is minimal.}
\label{opt}
\end{figure}

To examine the impact of link-targeted attack on the evolution of cooperation in
PGGs more precisely, we show in Fig.~\ref{reward}(b) the dependence of $\rho_{C}$ on $m_{L}$ for different values of $r$. It can be observed that link deletions have a predominantly negative impact on the evolution of cooperation. More specifically, for low values of $r$ the decrease of cooperation is most severe, which is in agreement with the fact that low multiplication factors are most challenging for the evolution of cooperation, and accordingly, the decrease in the network heterogeneity due to link deletions is there most notable. It can also be observed that for low values of $r$ there in fact exists an ``optimal'' number of deleted links at which defection thrives best (cooperators are most decimated). On the other hand, for larger $r$ the effect of link deletions is monotonously negative, i.e., the more links are deleted the smaller the level of cooperation on the network. For $r$ exceeding a threshold ($r>3.74$ \cite{szolnoki_pre09c}, which is linked to the survivability of cooperators on the square lattice), however, the negative impact of link deletions practically vanishes. This is because at such high values of $r$ the synergetic effects of cooperation suffice to withstand defector attacks even in the complete absence of network heterogeneity.

The reported decrease of cooperation upon link-targeted attack is consistent with previous results in that the connections among hubs facilitate cooperation in two-player networked games, such as the prisoner's dilemma or the snowdrift game \cite{santosXprl05, percXnjp09}, as well as in multi-player PGGs \cite{santosXn08}. In particular, a defector hub is frequently weak due to a negative feedback effect caused by its defecting neighborhood. Since there is then nobody left to exploit, a cooperator linked to such a defecting hub can overtake it and thus recover cooperation in a notable portion of the network.

\begin{figure}
\centerline{\epsfig{file=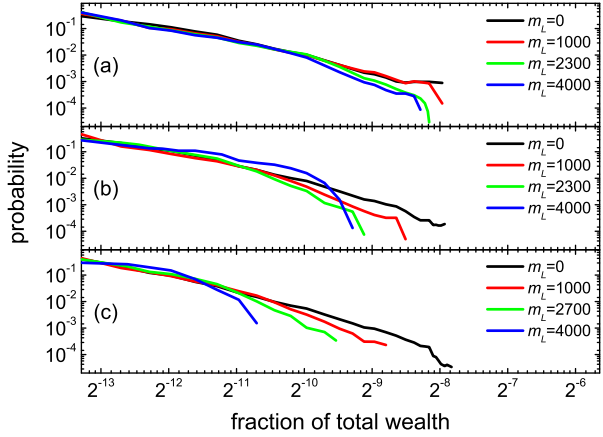,width=8.0cm}} \caption{Wealth distributions resulting from multi-group PGGs for different numbers of deleted links $m_L$ connected to the highest degree players. Values of the multiplication factor are: (a) $r=2.5$, (b) $r=3.0$ and (c) $r=4.0$.}
\label{mpay}
\end{figure}

Turning to results presented in Fig.~\ref{reward}(b), we elaborate further on the existence of an optimal number of link deletions, denoted here by $m_{L}^{opt}$, at which defectors are most widespread. Results presented in Fig.~\ref{opt} indicate that there exists only a rather narrow range of $r$ values (hatched gray), at which the link deletions have a monotonously negative impact on the evolution of cooperation. Outside this region, however, there always exists a non-zero value of $m_{L}^{opt}$ at which $\rho_C$ is minimal. We note, nevertheless, that following what is usually a sharp decrease of $\rho_{C}$ [especially for low values of $r$; see Fig.~\ref{reward}(b)], the subsequent increase of $\rho_{C}$ upon increasing $m_{L}$ past $m_{L}^{opt}$ is very modest indeed. The effect is mainly due to the creation of very isolated (although not fully disconnected) groups of cooperators that can prevail against defectors due to their isolation. This can be interpreted as group selection, although in our case the creation of groups is a sole consequence of the sparse network structure that emergences when $m_L$ approaches the transition point for network disintegration (see also results presented in Fig.~\ref{networks}).

Finally, to gain further insights, we examine the resulting wealth distributions for different values of $m_L$ and $r$, as presented in Fig.~\ref{mpay}. It can be observed that multi-group PGGs induce heterogeneous wealth distributions that are characterized by a fat tail. The algebraic wealth distribution agrees well with previous results \cite{santosXn08, YWWLW:2009, HWJWW:2006}. Notably, as the number of removed links increases, there emerges a cutoff in the distribution for the largest total wealth. The resulting deviation from the scale-free distribution obtained for $m_L=0$ induces a downfall of cooperation, especially for small values of $r$, as evidenced in Fig.~\ref{reward}(b). It is next of interest to summon our findings reported thus far and provide an analytical treatment to corroborate them.

\section{Analytical treatment}

In what follows, we predict the payoffs of agents with respect to their degrees as
well as the corresponding wealth distribution by virtue of an analytical approach. Assuming that cooperators are randomly distributed on the network and the fraction of cooperators is given by $\rho_C$, we can express the
payoff $p_{i,j}$ as
\begin{equation} \label{eq:mxy}
p_{i,j} = \frac{r\rho_C}{k_j +1}\left
(\sum_{l=1}^{N}\frac{A_{jl}}{k_l+1} +\frac{1}{k_j + 1} \right)-
\frac{\rho_C}{k_i +1},
\end{equation}
where $A_{jl}$ is the adjacency matrix of the underlying network.
Neglecting the degree-degree correlation, we have
\begin{eqnarray} \label{eq:average_A}
\sum_{l=1}^{N}\frac{A_{jl}}{k_l+1} &=& k_j\sum_{k_{min}}^{k_{max}}
\frac{\bar{P}(k'|k_j)}{k' +1} = k_j\sum_{k_{min}}^{k_{max}}
\frac{k' \bar{P}(k')}{\langle k\rangle} \frac{1}{k' +1}\nonumber \\
&=& k_y \left\langle \frac{k}{k+1} \right\rangle /\langle
k\rangle,
\end{eqnarray}
where $\langle \cdot \rangle$ denotes the average of all nodes,
$P(k')$ is the degree distribution of network nodes,
$\bar{P}(k'|k_j)$ is the joint degree distribution defined by
$\bar{P}(k'|k_j) = k' \bar{P}(k')/\langle k\rangle$, and the
identity $ \sum_{k_{min}}^{k_{max}} P(k') k'/(k' +1) = \left
\langle k/(k+1) \right \rangle $ has been used. Substituting Eq.
(\ref{eq:average_A}) in Eq. (\ref{eq:mxy}), we obtain
\begin{equation}
p_{i,j} =\rho_C \left[ \frac{r}{\langle k\rangle} \left\langle
\frac{k}{k+1} \right\rangle \frac{ k_j}{k_j +1} + \frac{r}{(k_j +
1)^2} - \frac{1}{k_i +1} \right].
\end{equation}
The total payoff is then given by
\begin{eqnarray}
&P_i& = \sum_{j=1}^{N}A_{ij} p_{i,j} + p_{i,i} \nonumber
\\&=& \rho_C
\bigg{[}\frac{r}{\langle k\rangle^2} \left\langle
\frac{k}{k+1}\right\rangle \left\langle
\frac{k^2}{k+1}\right\rangle  k_i + \frac{r}{\langle k\rangle}
\left \langle \frac{k}{(k+1)^2} \right\rangle k_i \nonumber \\&&+
\frac{r}{\langle k\rangle} \left\langle \frac{k}{k+1}
\right\rangle \frac{ k_i}{k_i +1} + \frac{r}{(k_i + 1)^2}
-1\bigg{]}. \label{eq:payoff_2}
\end{eqnarray}
For $k_i \gg 1$, we can obtain
\begin{equation}
P_i \approx Ak_i + B,
\end{equation}
where
\begin{eqnarray}
A &=& \rho_C \bigg{[}\frac{r}{\langle k\rangle^2} \left\langle
\frac{k}{k+1}\right\rangle \left\langle
\frac{k^2}{k+1}\right\rangle  + \frac{r}{\langle k\rangle}
\left \langle \frac{k}{(k+1)^2} \right\rangle \bigg{]}, \nonumber \\
B &=& \rho_C \bigg{[}\frac{r}{\langle k\rangle} \left\langle
\frac{k}{k+1} \right\rangle  -1\bigg{]}.
\end{eqnarray}
Here we can have $A \gtrsim B$, $B > -1$ and $k_i \gg 1$, $P_i$ is
thus proportional to $k_i$. Distribution of wealth $\bar{P}(P)$
can then be obtained by $\bar{P}(k)dk = \bar{P}(P)dP$ for
$\bar{P}(k)=\langle k\rangle^2/2 k^{-3}$:
\begin{equation}
\bar{P}(P)=\bar{P}(k)\frac{dk}{dP} = \frac{A^2 \langle k
\rangle^2}{2}(P-B)^{-3}.
\end{equation}

Since in the mean-field approximation the payoffs of hubs are proportional to their degrees, the connections among hubs play a key role in the maintenance of cooperation \cite{santosXprl05, santosXn08, percXnjp09}. The deletion of links among hubs thus
prevents cooperation, as shown in Fig.~\ref{reward}(b).

\section{Summary}

We have investigated the evolution of cooperation in multi-group PGGs on scale-free networks that were subject to link-targeted intentional attack. The main finding is that the deletion of links connected to hubs has a predominantly negative impact on the evolution of cooperation when players can participate in multiple groups. However, the effect is strictly negative (the more links are removed the lower the fraction of cooperators) only in a rather narrow region of the multiplication factor, whereas outside this region there exists an intermediate number of removed links at which defection thrives best. The latter effect, however, is quite subtle and is due to the emergence of group selection that sets in because of the sparse network structure that occurs when large numbers of links are removed. The results can be explained by means of wealth distributions and an analytical treatment that is further supported by heuristic arguments, albeit the subtleties of the existence of an intermediate number of link removals for highest levels of defection cannot be precisely accounted for (due to the emergence of a rather contrived version of group selection, as described). To sum up, we confirm that socially diverse states are crucial for the successful evolution of cooperation, where the upper limit of personal wealth introduces a cutoff that warrants the most challenging environment for the spread of defection. Given these facts, strong parallels can be drawn between the evolution of cooperation on complex networks and processes such as spread of viral diseases or the effectiveness of information retrieval, which were also found to be strongly affected by intentional attack \cite{newmanXsiam03}. We hope that our work will inspire further research in this direction and that it will serve as a useful source of information when striving towards the understanding of the stability of cooperation under attack in social networks.

\begin{acknowledgments}
LLJ and BHW were supported by the National Natural Science Foundation of China (Grant Nos. 11047012 and 10975126). MP was supported by the Slovenian Research Agency (Grant No. Z1-2032). WXW and YCL were supported by AFOSR (Grant No. FA9550-10-1-0083).
\end{acknowledgments}

\end{document}